\documentclass{emulateapj}

\def\lya{\ifmmode {\rm Ly}\alpha~ \else Ly$\alpha$\fi}
\def\ci{\ifmmode {\rm C}\,{\sc i}~ \else C\,{\sc i}\fi}
\def\ciii{\ifmmode {\rm C}\,{\sc iii}~ \else C\,{\sc iii}\fi}
\def\civ{\ifmmode {\rm C}\,{\sc iv}~ \else C\,{\sc iv}\fi}
\def\civs{\ifmmode {\rm C}\,{\sc iv}~ \else C\,{\sc iv}~\fi}
\def\nv{\ifmmode {\rm N}\,{\sc v}~ \else N\,{\sc v}\fi}
\def\nvs{\ifmmode {\rm N}\,{\sc v}~ \else N\,{\sc v}~\fi}
\def\alii{Al\,{\sc ii}}
\def\siii{Si\,{\sc ii}}

\def\siivs{Si\,{\sc iv}~}
\def\oi{\ifmmode {\rm O}\,{\sc i} \else O\,{\sc i}\fi}

\def\ovi{O\,{\sc vi}}

\def\lyb{Ly$\beta$}

\def\msun{\ifmmode {\rm ~M}_\odot~ \else M$_\odot$\fi}

\def\5100{$\lambda L_\lambda$(5100 \AA)}

\def\LLedd{$L/L_{Edd}$}
\shorttitle{Scaling Ultraviolet Outflows in Seyferts}
\shortauthors{Stoll et al.}

\begin{document}

\title{Scaling Ultraviolet Outflows in Seyferts}

\author{R. Stoll,\altaffilmark{1} S. Mathur,\altaffilmark{1} Y. Krongold,\altaffilmark{2} and F. Nicastro\altaffilmark{3}}
\email{stoll@astronomy.ohio-state.edu}

\altaffiltext{1}{Department of Astronomy, The Ohio State University, Columbus, OH 43210}
\altaffiltext{2}{Instituto de Astronomia, Universidad Nacional Autonoma de Mexico, Apartado Postal 70-264, 04510 Mexico DF, Mexico}
\altaffiltext{3}{Osservatorio Astronomico di Roma (INAF), Via Frascati 33, Rome I-00040 Italy}

\begin{abstract}

X-ray and UV absorbing outflows are frequently seen in AGN and have been cited 
as a possible feedback mechanism.  Whether or not they can provide adequate 
feedback depends on how massive they are and how much energy they carry, 
but it depends in a more fundamental way upon whether they escape the 
potential of the black hole.  If the outflows have reached their asymptotic 
velocity when we observe them, then all of these properties critically depend 
on the radius of the outflow: a value which is difficult to measure.  The 
tightest limit on the distance of an X-ray warm absorber from the ionizing 
source is that of \citet{krongold2007} for NGC~4051.  We use NGC~4051 to model 
other observed UV outflows, and find that on the whole they may not provide 
meaningful feedback.  The outflow velocities are below or just above the escape 
velocity of the black hole.  This may be because they are not yet fully 
accelerated, or the duty cycle of high-velocity outflows may be small.  
Another possibility is that they may only provide meaningful feedback in 
higher-luminosity AGN, as we find a weak correlation between the ratio of 
outflow velocity to escape velocity and AGN continuum luminosity.  

\end{abstract}


\section{Introduction}\label{intro}
A large fraction of AGN show signs of circumnuclear outflows, which manifest 
themselves as UV absorption lines and as X-ray ``warm absorbers'' 
\citep[and references therein]{reynolds1997,george1998,crenshawARAA}.  
Understanding the kinematics and dynamics of these outflows and their 
detailed physical conditions is important for a complete understanding of the 
nuclear region of AGN \citep[e.g.][]{mathur1994,krolik1995,mathur1995,elvis2000}.  
Theoretical models such as that of \citet{proga2007} make specific 
predictions about the properties of the outflow and their relation to 
the properties of the AGN.  Direct comparison between the models and 
observations is still lacking, however, because measuring the mass and energy 
outflow rates is a difficult problem.

Additional impetus for studying the outflows comes from the fact that they
provide a possible mechanism for feedback.  Feedback from AGN has been 
invoked to attempt to solve a number of astrophysical problems, from 
cluster cooling flows to the structure of galaxies.  Feedback from jets 
appears to be sufficient to keep the cooling flows in clusters from cooling
too much \citep[e.g.][]{mcnamara2001,oh2003}, and may be sufficient to 
regulate black hole (BH) growth in central cluster galaxies 
\citep[e.g.][]{rafferty2006}.  Only about 10\% of all quasars are radio-loud,
however \citep{white2007}, so either feedback from powerful radio jets 
cannot be universal or the duty cycle is small.  
Circumnuclear outflows could potentially be a more common form of quasar 
feedback, as intrinsic absorption appears in approximately 60\% of all
AGN \citep{ganguly2008}.

We know from absorption and emission-line studies \citep[e.g.][]{fields2007}
that high-metallicity gas is present near galactic nuclei, so these outflows
might also be responsible for enriching the intergalactic medium with metals.
In order to do so, the outflows must escape the global
gravitational potential and must entrain this high-metallicity gas.
Understanding the physical conditions in the absorbers thus becomes
very important, particularly their energy outflow rates and their masses.

While non-jet feedback has proved important in theoretical models, we do not 
really know if it works or really occurs.
In theoretical examinations of intragalactic AGN feedback, 
energy injection efficiencies range from the order of unity
($L_{outflow}/L_{bolometric} \approx 1$) to a minimum of 5\% 
\citep[e.g.~][]{silk2005,scan2004}.  
Do we know whether actual AGN UV/X-ray warm absorber outflows indeed carry 
such energy, if naively considered as already at their asymptotic velocity?  
This is a challenging measurement to make, one which is critically dependent 
upon finding the radius of the absorber.

Determining the mass and energy of an outflow (and the feedback that it 
naively can provide) begins with the properties directly 
observable from absorption-line studies: column densities for a variety 
of ions.  Photoionization modeling allows us to convert the ionic column 
densities to the column densities of metals.  To go from column density to 
mass requires knowing the metallicity of the gas and the location and 
geometry of the absorbing medium.  In general, data on associated absorption
systems in Seyferts have not been sensitive to metallicity, but 
super-solar metallicity absorbing gas has recently been discovered in two 
Seyferts \citep{fields2005,fields2007}.  
Inferred energy and mass loss rates depend strongly on the geometry of 
the absorber.  If it is a shell of gas located far from the nuclear 
black hole, for example, then the implied mass can be quite large 
compared with an absorber located closer in for a given column density.  
Once the geometry has been established (assumed or observed), 
the energy outflow rate can then be calculated from the mass and velocity 
(measured from the blueshifts of lines). It is vital to know the location 
of the absorber to reliably measure the amount of energy that feedback can 
inject into the surrounding medium.  The distance of the absorbing region 
($R$) from the ionizing source, however, is degenerate with the 
density ($n_e$) in the equation for the photoionization parameter 
($U \propto L/n_e R^2$).  This degeneracy can be broken if we can determine
the density independently.  Since the recombination times are inversely 
proportional to density, the response of absorption lines to continuum 
variations during the ionizing phase provides a robust density diagnostic.
We must therefore probe the appropriate time domain for an ionizing/recombining 
(rather than photoionization-equilibrium) plasma.

This observationally challenging technique has mostly produced upper limits
on the distances of outflows from the source (which can be reasonably argued
to be closely related to upper limits on the radii of the outflows) but it
has been successfully employed to calculate outflow mass and energy rates for 
the high ionization parameter component in NGC~4051 \citep{krongold2007}.  
We concentrate on the 
low ionization parameter component, for which \citet{krongold2007} also 
established a tighter limit, because it is associated with the UV absorption.  
The outflow distance for this component is small, at most $8.9 \times 10^{15}$ 
cm or 3.4 light-days.  The motivation behind this paper is their surprising 
discovery that the observed outflow rates are four to five orders of magnitude 
below those naively required for efficient feedback.  The X-ray outflow 
velocity in NGC~4051 is only 490 km s$^{-1}$, a small fraction of the escape 
velocity at its location.  
Can we generalize this result to all AGN, or even just to Seyferts?  After 
all, NGC~4051 is an unusual object, a narrow-line Seyfert 1 galaxy with low 
luminosity and a low-mass black hole, and it is highly variable.  Perhaps in 
other AGN the outflow velocity ($v_{out}$) is larger than the escape velocity 
($v_{esc}$), allowing feedback to occur even if the outflows are at their 
asymptotic velocities.  To test whether this is indeed the case, we compared 
the two in a number of AGN.  We discuss our method in \S \ref{data}, our 
results in \S \ref{results}, and we conclude in \S \ref{concl}.

\section{Data and Methods}\label{data}

In order to compare the ratio $v_{out}/v_{esc}$ of NGC~4051 to other AGN, 
we should ideally consider the velocity of the X-ray outflow, as 
\citet{krongold2007} do.  However, there are not very many high-resolution
grating X-ray observations which have measured the outflow velocity, 
$v_{out}$, so we have compiled $v_{out}$ as measured in UV outflows.  
The X-ray $v_{out}$ found by \citet{krongold2007} for NGC~4051 is 
492 $\pm$ 17 km s$^{-1}$, while it shows UV  outflows with $v_{out}$ ranging
from 48 to 727  km s$^{-1}$, and a median of around 360
\citep{collinge2001,dunn2008}.  
The UV and X-ray outflow velocities in NGC~4051 are the same order of 
magnitude; to the extent that they are different, the median UV 
outflow velocity slightly underestimates the outflow energy, but the 
difference is small.
It should be noted that a consistent picture of the nature of the highly 
ionized X-ray absorbers has emerged from recent observations of several AGN.  
The absorbers have at least two components: one with a low ionization 
parameter (LIP) and one with a high ionization parameter (HIP).  
The LIP and HIP phases appear to be 
in pressure equilibrium, and so likely emerge from the common wind 
\citep[e.g.][]{netzer2003,krongold2003}.  
The LIP component is also responsible for the UV absorption lines, which often
show multiple components unresolved in X-ray, but the HIP is not.
The UV outflow velocity is then essentially the $v_{out}$ of the LIP
component of the X-ray outflow.

The outflow velocities we use are from the literature, derived from intrinsic 
absorption lines in the near and far ultraviolet, including  
\lya, \lyb, \nv, \ci, \ciii, \civ, \oi, \ovi, \alii, \siii, and \siivs
\citep{brotherton2002,collinge2001,crenshaw1999,crenshaw2001,crenshaw2002,crenshaw2003,dunn2008,gabel2003a,gabel2005,kraemer2001a,kraemer2001b,kraemer2002,kraemer2003,kriss2000,kriss2003,romano2002}.  
See Table \ref{reftable}.
The spectra were taken with the Goddard High-Resolution Spectrograph (GHRS), 
the Faint Object Spectrograph (FOS), 
the Space Telescope Imaging Spectrograph (STIS) on HST, 
or with the Far Ultraviolet Spectroscopic Explorer (FUSE). 
We exclude objects for which we do not have reasonable black hole mass 
estimates (ESO265-G23, RX J1230.8+0115, and TOL1238-368).

\begin{deluxetable*}{llllll}
\tabletypesize{\footnotesize}
\tablecolumns{6}
\tablecaption{AGN parameters from the literature}
\tablehead{\colhead{log \5100}                                 &
           \colhead{$\sigma_L/L$}                              &
           \colhead{Object}                                    &
           \colhead{$M$}                                       &
           \colhead{$\sigma_M/M$}                              &
           \colhead{Ref ($M_{BH}$\tablenotemark{a}; $\lambda L_\lambda$\tablenotemark{b}; $v_{out}$\tablenotemark{c})}         
}
\startdata
43.59     &  .25    &  WPVS 007         & 1.15 $\times 10^7$    &  1.4   &   2;  C;  3,7        \\
44.79     &  .22    &  I Zw 1           & 2.76 $\times 10^7$    &  1.4   &   7;  F;  3          \\
44.27     &  1.5    &  TON S180         & 1.02 $\times 10^7$    &  1.4   &   4;  D;  7          \\
43.16     &  1.5    &  Mrk 1044         & 1.7  $\times 10^7$    &  1.2   &   8;  D;  7          \\
43.80     &  1.5    &  NGC 985          & 2.03 $\times 10^8$    &  1.4   &   4;  D;  7          \\
44.87     &  1.5    &  IRAS F04250-5718 & 1.44 $\times 10^8$    &  1.4   &   4;  D;  7          \\
43.65     & .0069   &  Mrk 79           & 5.24 $\times 10^7$    & 0.27   &   6;  A;  7          \\
43.08     &  .3     &  Mrk 10           & 2.56 $\times 10^7$    &  1.4   &   2;  B;  7          \\
45.39     &  .18    &  IRAS F07546+3928 & 1.77 $\times 10^8$    &  1.4   &   5;  E;  7          \\
42.48     &  .10    &  NGC 3227         & 4.22 $\times 10^7$    & 0.51   &   6;  A;  4          \\
42.62     &  .57    &  NGC 3516         & 4.27 $\times 10^7$    & 0.34   &   6;  A;  3,7,12     \\
43.02     &  .14    &  NGC 3783         & 2.98 $\times 10^7$    & 0.18   &   6;  A;  3,7,8,10   \\
41.88     &  .18    &  NGC 4051         & 1.91 $\times 10^6$    & 0.41   &   6;  A;  2,7        \\
41.92     &  .49    &  NGC 4151         & 4.57 $\times 10^7$    &  1.3   &   1;  A;  3,6,7,11   \\
44.84     &  .03    &  PG 1351+640      & 6.73 $\times 10^8$    &  1.4   &   7;  F;  7          \\
43.66     &  .18    &  Mrk 279          & 3.49 $\times 10^7$    & 0.26   &   6;  A;  7,9        \\
43.98     &  1.5    &  RX J1355.2+5612  & 9.4  $\times 10^6$    &  1.4   &   2;  D;  7          \\
44.38     &  .039   &  PG 1404+226      & 7.75 $\times 10^6$    &  1.4   &   7;  F;  7          \\
43.31     &  .054   &  NGC 5548         & 6.71 $\times 10^7$    & .039   &   6;  A;  1,3,6      \\
43.64     &  .064   &  Mrk 817          & 4.94 $\times 10^7$    & 0.16   &   6;  A;  7          \\
43.69     &  .023   &  Mrk 290          & 1.60 $\times 10^8$    &  1.4   &   7;  F;  7          \\
44.73     &  .22    &  Mrk 876          & 2.79 $\times 10^8$    & 0.46   &   6;  A;  7          \\
44.16     &  .23    &  Mrk 509          & 1.43 $\times 10^8$    & .084   &   6;  A;  3,7,13,14  \\
44.40     &  .056   &  II Zw 136        & 3.8  $\times 10^7$    & 0.39   &   3;  A;  3,7        \\
43.62     &  .3     &  Akn 564          & 4.8  $\times 10^6$    &  1.4   &   2;  B;  3,5,7,16   \\
42.76     &  .23    &  IRAS F22456-5125 & 1.55 $\times 10^8$    &  1.4   &   2;  C;  7          \\
44.64     &  .092   &  MR 2251-178      & 2.40 $\times 10^8$    &  1.4   &   2;  C;  7          \\
43.30     &  .12    &  NGC 7469         & 1.22 $\times 10^7$    & 0.11   &   6;  A;  3,7,15     \\
\enddata

\tablenotetext{a}{References for $M_{BH}$. 
1:~\citet{bentz2006b}, 
2:~\citet{dunn2008},
3:~\citet{grier2008},
4:~\citet{grupe2004},
5:~\citet{marziani2003},
6:~\citet{peterson2004}, 
7:~\citet{vestergaard2006},
8:~\citet{watson2007}
}

\tablenotetext{b}{References for \5100.
A:~\citet{bentz2008},
B:~\citet{botte2004},
C:~\citet{dunn2008},
D:~\citet{grupe2004},
E:~\citet{marziani2003},
F:~\citet{vestergaard2006}
}

\tablenotetext{c}{References for $v_{outflow}$.
1:~\citet{brotherton2002},
2:~\citet{collinge2001},
3:~\citet{crenshaw1999},
4:~\citet{crenshaw2001},
5:~\citet{crenshaw2002},
6:~\citet{crenshaw2003},
7:~\citet{dunn2008},
8:~\citet{gabel2003a},
9:~\citet{gabel2005},
10:~\citet{kraemer2001a},
11:~\citet{kraemer2001b},
12:~\citet{kraemer2002},
13:~\citet{kraemer2003},
14:~\citet{kriss2000},
15:~\citet{kriss2003},
16:~\citet{romano2002}
}

\label{reftable}

\end{deluxetable*}


In order to calculate the escape velocity at the wind-launching radius, we need
to estimate the radius.  As discussed above, the distance of the absorber from 
the ionizing source has not been reliably measured for any AGN other than 
NGC~4051, so we scale the absorber radius of the other AGN from the NGC~4051 
distance.  We use two different scalings.

The first of these is 
\begin{equation}\label{Lscaling} R \propto \sqrt{\lambda L_{\lambda}}, \end{equation} 
the relationship demonstrated by \citet{bentz2006a,bentz2008} for the radius 
of the broad-line region.  It is logical to extend this essentially geometrical 
relationship to other radii which are defined by photoionization, like that 
of the UV/X-ray warm absorbers.  We used 5100~\AA~starlight-subtracted 
continuum luminosity from \citet{bentz2008} or \citet{vestergaard2006}
when available.  When not, we used luminosities from \citet{grupe2004},
\citet{botte2004}, and \citet{dunn2008}.

For the second of these scalings,  we try an alternative, gravitational
scaling, again using the radius of the warm absorber of NGC~4051 measured by
\citet{krongold2007} and scaling linearly by the black hole mass.  
\begin{equation}\label{Mscaling} R \propto M_{BH} \end{equation}
This scaling is not motivated by theory; it is merely an attempt to scale to 
one of the few remaining fundamental parameters of the system.

To calculate the escape velocity at the scaled radius of the absorber, 
we used reverberation mapping rms black hole masses 
\citep{grier2008,peterson2004,bentz2006b} when available, then
masses from single-epoch scaling relations calculated with 
starlight-corrected continuum luminosities when possible 
\citep{vestergaard2006}.  We also use single-epoch scaling relations from 
\citet{dunn2008}, \citet{grupe2004}, \citet{marziani2003}, and 
\citet{watson2007}.

\section{Results and Discussion}\label{results}
\begin{figure}
\plotone{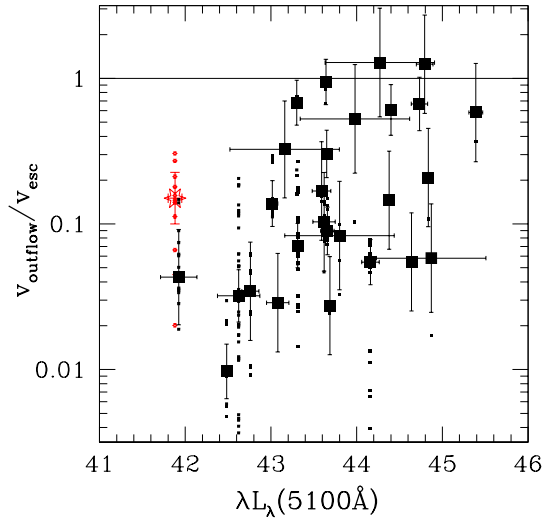}
\caption{Outflow velocity in units of escape velocity plotted against AGN 
luminosity.  All outflow radii are scaled by $\sqrt{\lambda L_{\lambda}}$
to the X-ray absorber in NGC~4051 for which \citet{krongold2007} 
measured the radius, marked with a star.  For objects with intrinsic UV
absorption at several velocities, the outflow with median velocity is 
plotted as a large square and the others are plotted as small squares.
\label{figL}}
\end{figure}

In Figure \ref{figL} we plot the ratio of the outflow velocity 
to the escape velocity determined from the first scaling relationship 
(eqn.~\ref{Lscaling}).  
The large open star marks the median UV outflow velocity of 
NGC~4051.\footnote{This is a larger fraction of the escape velocity than is 
discussed in \citet{krongold2007}.  They focus on the HIP, while we are 
primarily concerned with the LIP, which produces the UV lines.  
}
The smaller symbols are the other observed (line-of-sight) outflow velocities.  
The outflow velocity ratios for the other objects, which we obtained by 
scaling, are plotted as squares; again the median $v_{out}$ for each object 
is plotted with a larger symbol.
It is interesting to note that for more than two thirds of the AGN, the 
maximum outflow velocity is less than half the escape velocity.  Assuming 
that the (unknown) transverse velocities of the outflows are similar to 
their line-of-sight velocities, this would imply that the outflows in over 
two-thirds of these objects are moving at less than the escape velocity.
We find a weak (0.42) correlation between the median values of
$\log (v_{out}/v_{esc})$ and $\log \lambda L_\lambda$, with a 2\% probability of
appearing by chance.  (This and all correlation coefficients in this paper 
are Spearman's rank.)  The similar correlation for the maximum observed
outflow velocity in each AGN is much weaker, and is consistent with statistical 
fluctuations.  
This correlation may imply 
that feedback is more efficient for higher-luminosity AGN, but it may simply 
mean that since the location for which we observe the UV outflows in 
higher-luminosity AGN is farther away from the black hole, they have been more 
highly accelerated before reaching the point at which we observe them.  
Because it is a weak correlation, it is likely that if there is real physics 
here, we are observing a secondary rather than a primary effect.
Some of the spread may be due to the fact that we use the 
published intrinsic absorption velocities from whatever epochs are available, 
but we only use a single continuum luminosity value for each object, because in 
general \5100 measurements are not published for the same 
epochs, much less the starlight-subtracted values we favor for these 
low-luminosity AGN.  It would be interesting to see how the outflow velocities 
vary with continuum luminosity changes for a single object, and compare this
to expectations for radiatively- or thermally-driven winds.

\begin{figure}
\plotone{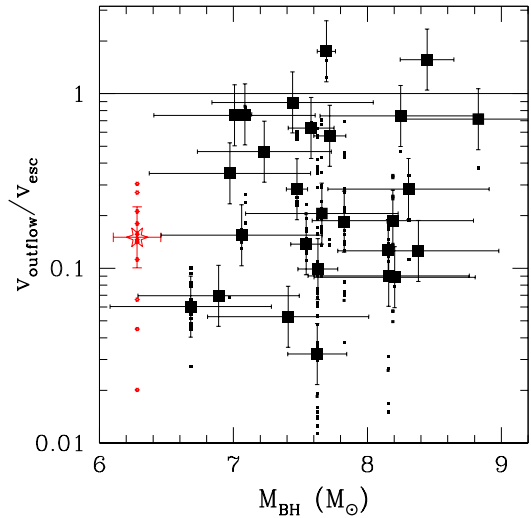}
\caption{Outflow velocity in units of escape velocity plotted against black 
hole mass.  All outflow radii are scaled by $M_{BH}$ to NGC~4051, 
marked with a star.  Points as in Figure \ref{figL}.
\label{figM}}
\end{figure}

In Figure \ref{figM} the same velocity ratio is plotted, but with the wind 
radius estimated using the mass scaling relationship (eqn. \ref{Mscaling}).  
Again, we see that for all but two AGN the observed $v_{outflow} < v_{esc}$.  
The general conclusion that the outflow velocities for these Seyferts are 
generally a small fraction of the escape velocities appears to be independent 
of the scaling relation used.  There is no significant trend for this scaling 
relation with the black hole mass (Fig. \ref{figM}), and there is a very weak 
(0.32) correlation with the luminosity that is marginally statistically 
significant (10\% probability of chance), which we do not plot. 

\begin{figure}
\plotone{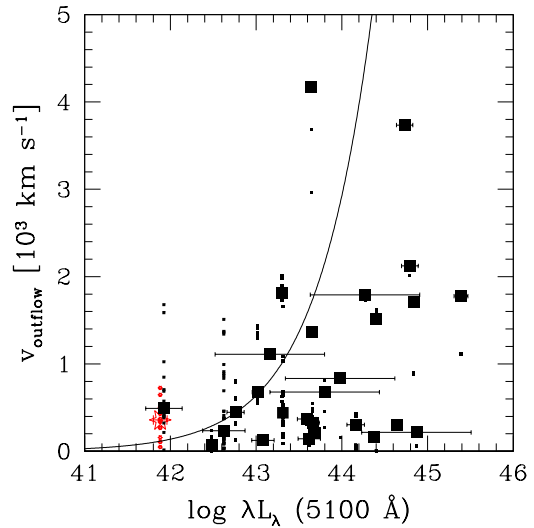}
\caption{Outflow velocity plotted against AGN luminosity.  
Line is from \citet{ganguly2007}, a refit of the envelope found by 
\citet{laor2002}.  Note that NGC~4051 (again, marked with a star) lies 
significantly outside the envelope.  None of the information in this figure
is scaled to NGC~4051.
\label{figenv}}
\end{figure}

How do the data from which we extrapolate compare with other studies in the 
literature?  The kinematics of UV outflows in AGN has been investigated 
by \citet{laor2002}
and more recently by \citet{ganguly2008} and \citet{dunn2008}.
These studies find that there is an upper envelope to the relation between 
the maximum velocity of outflows and the AGN luminosity.  The two are directly
correlated only for the soft-X-ray weak quasars (SXWQs), which trace the 
envelope \citep{laor2002}.   
The upper envelope of the trend is described by  $v_{max} \propto L^\alpha$
where $\alpha = 0.662 \pm 0.004$ \citep[who refit the SXWQs of \citet{laor2002} 
updating to a standard $\Lambda$CDM cosmology]{ganguly2007}.  We overplot
this envelope on the data from which we extrapolate in Figure \ref{figenv}.  
Notice that outflows for some of the AGN, including NGC~4051 (plotted with an 
open star) lie above the relation at this normalization, most probably because 
the AGN continuum luminosities we use are corrected for starlight 
contamination to better reflect the properties of the AGN itself.  We do see 
a general consistency with the previously observed relationship.

\subsection{Implications for the nature of outflows and the feedback mechanism}

Taken at face value, Figures \ref{figL} and \ref{figM} imply that the outflow
velocity in most AGN is so small compared with the escape velocity that the 
outflow may never leave the circumnuclear region of the AGN unless it is
still being accelerated when we observe it.  UV inflows (blueshifted intrinsic 
absorption lines) are very rarely observed in AGN, suggesting either that all
such outflows do indeed escape or that they change in characteristics before
falling in, perhaps becoming less dense or clumpy.  Radiatively-driven outflows 
in an optically thin environment ought to continue to accelerate asymptotically 
to a final velocity $v_\infty$, because both the gravity and the 
radiation-pressure force decrease as $1/r^2$.  

\begin{figure}
\plotone{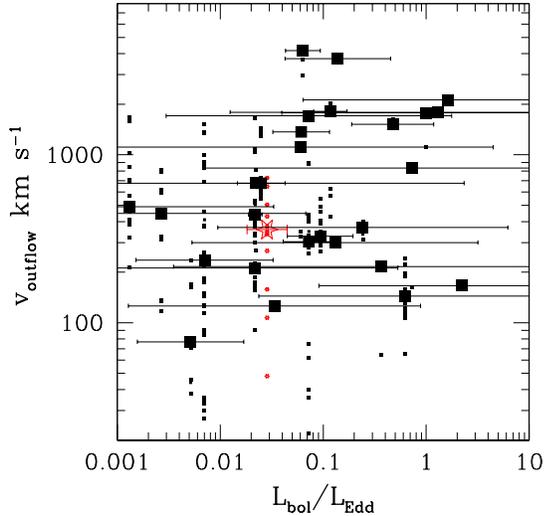}
\caption{Same scaline and plotting convention as figures \ref{figL} and 
\ref{figenv}, but plotted to show the lack of correlation between outflow
velocity and Eddington ratio, assuming $L_{bol} = 9$\5100.
\label{figLedd}}
\end{figure}

\begin{figure}
\plotone{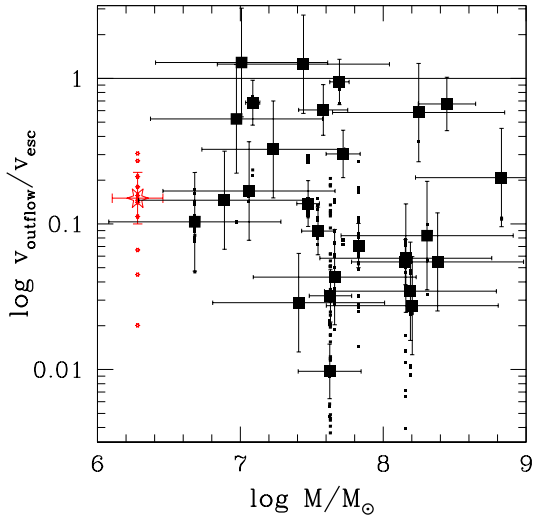}
\caption{Outflow velocity ratios with escape velocity scaled with our
primary scaling relationship ($\propto \sqrt{L}$) but plotted against
mass.  Compare with feedback models such as that of \citet{scan2004},
which implement feedback as a fraction of the luminosity of an AGN
emitting at Eddington.  We find no mass-dependence with this scaling
to support a model like that.
\label{figLsclM}}
\end{figure}

The disk-wind models of \citet{proga2007} have shown that efficient mass and
energy outflow results from radiation-driving for a $10^8$ M$_\cdot$ black hole 
at L/L$_{Edd}$ = 0.6.  The efficiency of these radiatively-driven outflows in 
turn depends upon black hole mass ($M_{BH}$) and the accretion rate relative to 
the Eddington limit ($L/L_{Edd}$) \citep{proga2004,proga2005}.  
We see no clear dependence of $v_{out}/v_{esc}$ on 
$M_{BH}$ with either scaling relation (see Figs. \ref{figM} and 
\ref{figLsclM}).  It would be nice to see how $v_{out}/v_{esc}$ depends on 
\LLedd, but these two highly-derived quantities both depend on the mass of 
the black hole.  Any extra information to be gained is masked by the spurious 
correlation that is due to essentially plotting $\sigma_{line}^{-1}$ vs 
$\sigma_{line}^{-2}$, where $\sigma_{line}$ is e.g.~the width of the H$\beta$ 
broad line.  The best that can be done instead is to say that we see no 
dependence of $v_{out}$ on $L_{bol}/L_{Edd}$, as Fig. \ref{figLedd} shows, 
where we assume $L_{bol} \approx 9$\5100 \citep[as][]{kaspi2000} and use our 
geometrically motivated scaling relationship (eqn. \ref{Lscaling}).  Perhaps 
the gas is not accelerated as efficiently as in the models of \citet{proga2007}.

We know that high velocity outflows exist in broad absorption line quasars 
(BALQSOs), which make up approximately 17-40\% of all luminous quasars 
\citep{knigge2008,dai2008}.  (Estimates vary, due in part to quasar selection
criteria and in part to differing definitions of what constitutes a BALQSO 
\citep{weymann1991,trump2006,ganguly2008,knigge2008}).  Their highest 
velocities, up to about 0.4c \citep{chartas2002}, 
are so large that they reach escape 
velocity even if they arise in the inner parts of accretion disks. 
There may then be a qualitative difference between the properties of the 
outflows in the Seyfert galaxies in our sample and luminous BALQSOs.

\begin{figure}
\plotone{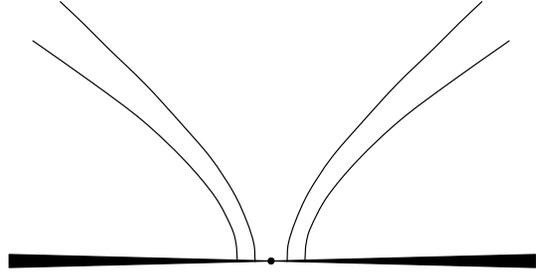}
\caption{Schematic diagram of an outflow geometry \citep[e.g.][]{elvis2000}; 
flare angle of winds shown with accretion disk.  If the flare angle is small
and if winds are not fully accelerated until the outer edge of the region 
then from most viewing angles the observed velocity of the winds will be 
misleadingly small.
\label{figWinds}}
\end{figure}

The geometry of the outflow 
may be such that we almost always observe the UV outflow before it is fully
accelerated.  This may happen, for example, if the opening angle of the 
biconical funnel \citep[e.g.][]{elvis2000} is small, making a view down the 
throat of the funnel highly unlikely (see Figure \ref{figWinds}).
If that is the case, then not only may we be observing them before they are
fully accelerated, but it also may be that there is always a large component
of transverse velocity in the UV outflows that we observe, though we only
measure line-of-sight velocity.  Limits on transverse velocities are 
generally of the same order as the radial velocities 
\citep[e.g.][]{crenshawARAA,gabel2003b}.
The total velocity of these outflows, then, may be high enough for them to 
provide meaningful feedback.  In such a geometry, however, the energy 
deposition would necessarily be highly anisotropic.  The same conceptual 
problem arises for jets, which also may provide only very localized energy 
deposition.  In both cases, the interaction between the outflow and the 
surrounding medium serves to isotropize the feedback to a certain extent.  
Can narrow outflows lead to effective feedback for the regulation of bulge 
growth and star formation in a galaxy?

A second possible way out is if neither of our two scaling relationships is
valid.  In that case we simply cannot extrapolate the nature of UV outflows
in AGN from the case of NGC~4051.  While this may be the case, it is not enough 
by itself to save the notion of effective feedback from UV outflows, because it 
begs the question of why neither a general geometric photoionization scaling 
nor a scaling based on mass is valid.  After all, the outflow models 
\citep[e.g.][]{hopkins2007,scan2004} do scale with bolometric 
luminosity.\footnote{In the model of \citet{scan2004}, feedback is implemented 
as a fraction of the luminosity, which is assumed to be Eddington until the AGN 
disrupts its fuel source, so in effect feedback in this model scales as the 
mass of the black hole, as in our Figure \ref{figLsclM}.}

If feedback is inefficient or does not occur for some AGN, then it has serious 
consequences for the feedback models of black hole growth and BH-galaxy 
coevolution \citep[e.g.][]{robertson2006,dimatteo2008}.  This in itself may 
not be a problem, as there are alternative feedback models which 
only invoke jet AGN activity.  
\citet{natarajan2008} invoke dark matter annihilation/decay as an
effective feedback mechanism that is entirely uncoupled to AGN activity.
\citet{croton2006} use only radio-mode
(jet) feedback, and yet they achieve most of the results of other feedback
models, including the galaxy luminosity function cutoff.  They do issue
the caveat, however, that non-jet feedback may be compensated for in their
model by enhanced stellar feedback.  If stellar feedback can indeed operate
at the efficiencies they invoke, then their model remains unaffected by the 
absence of effective intragalactic AGN feedback that our analysis suggests.

Perhaps the outflow-mode feedback works, and is energetic enough and isotropic 
enough, but does not manifest itself as a UV or X-ray warm absorber outflow.  
There have been several claims of relativistic outflows observed as highly 
shifted X-ray absorption or emission lines 
\citep[e.g.][]{turner2004,dadina2005,petrucci2007}.  Such outflows would 
certainly be energetic enough to escape and could potentially be effective as
feedback.  The existence of such relativistically shifted lines was questioned
recently by \citet{vaughan2008}, who argue that the reported outflows are 
consistent with statistical fluctuations, given publication bias.  
A viable outflow mechanism is not yet known, then, that can provide effective 
feedback for all AGN.

One more possibility is that UV/X-ray warm absorber outflows may provide 
meaningful feedback, but only with a very small duty cycle, and most of the 
time be observed only at the relatively low velocities we see in these Seyfert 
galaxies.  This sort of bursting behavior is superficially similar to the 
feedback patterns \citet{ciotti2009} are finding as they continue to refine 
their 1-D hydrodynamic feedback models.

Is it possible that this 
kind of feedback does not occur for Seyferts but does for more luminous AGN?
Perhaps the high-velocity outflows observed in some BALQSOs can provide 
feedback in some high-luminosity objects, but similar outflows can provide 
no feedback in low-luminosity objects (as our results suggest).  If this
were the case, and if AGN feedback is indeed necessary, then that would 
imply there must be some other form of feedback acting for these 
low-luminosity objects.  That two distinct mechanisms could form one 
smooth M-$\sigma$ relationship seems an unlikely conspiracy.  We are inclined
to reject the disjoint feedback mechanism hypothesis.

Another possibility is that feedback may be more efficient in higher-luminosity
AGN, as perhaps suggested by the weak correlation in our Figure \ref{figL}.  
This alone is not sufficient to solve the problem if the outflows are already 
at their asymptotic velocity when we observe them, however, because the most 
luminous quasars have bolometric luminosities around $10^{48}$ erg s$^{-1}$.  
Assuming again a bolometric correction of 9 to \5100, our objects extend four 
dex in luminosity, past $10^{46}$ erg s$^{-1}$, and span approximately two dex in 
$v_{out}/v_{esc}$.  With two more dex in luminosity, we would expect a 
$v_{out}/v_{esc}$ increase of about one dex, for a maximum $v_{out}/v_{esc}$ of 
around 10.  While these outflows would indeed escape the potential of the black 
hole, they would hardly escape far enough into the bulge with enough energy 
remaining to achieve the global-scale feedback that models invoke.

The final possibility to consider is that the UV outflows are experiencing 
continued acceleration when we observe them.  There is some evidence for this 
at a smaller scale in AGN; outflowing clumps in the narrow-line region of 
NGC 4151 seem to increase in velocity linearly with distance from the nucleus
\citep{das2005}.  Continued acceleration is expected for simple models of 
thermally or radiation-pressure-driven winds, though detailed modeling is 
difficult to reconcile with the observations \citep[e.g.][]{everett2007}.  
If the UV/X-ray warm absorbers are undergoing continued acceleration, efforts
like ours to localize them may be of some use in constraining wind models.

\section{Conclusions}\label{concl}

\citet{krongold2007} find that the UV/X-ray warm absorber outflow in NGC~4051
is insufficient to provide effective feedback to its host galaxy unless 
it is undergoing continued acceleration.  We investigated outflow velocities 
of other AGN but found them to fall significantly short of their escape 
velocities at the locations we extrapolate for them.  These results imply that 
either outflows do not provide a viable feedback mechanism or they are 
experiencing continued acceleration.
We have discussed caveats to these arguments, which lead to further questions.
Are the opening angles of the outflow too small to intercept edge-on, so that
we rarely observe their full velocities?  How will we test such a scenario?  
And if this is indeed the case, how do they deposit energy in an isotropic 
fashion to regulate the bulge mass?  If the outflows are experiencing 
continued acceleration, what is driving it?  Do the outflows provide 
significant feedback, but only with a very small duty cycle, so we do not 
observe them in these Seyferts?  Or do our results imply that AGN feedback 
in Seyferts just does not occur?  If feedback does not occur, what then is the 
origin of the $M_{BH}-\sigma$ relation?

In order to lay to rest the question of the efficacy of UV/X-ray warm absorber
outflows, it is crucial to measure the
radius of the outflow region in several AGN spanning a range of luminosities
and black hole masses.  Additional observations with XMM-Newton and the
upcoming Cosmic Origins Spectrograph (COS) on HST will be invaluable for
this purpose.

\acknowledgments
We thank Todd Thompson, Paul Martini, Molly Peeples, and Ramiro Sim\~{o}es Lopes
for valuable conversations.

\end{document}